\documentclass[pra,showpacs,twocolumn]{revtex4}
\usepackage{amsmath,amsfonts,graphicx}
\usepackage[amssymb]{SIunits}
\usepackage{color}

\newcommand{\ket}[1]{|#1\rangle}
\newcommand{\bra}[1]{\langle#1|}
\newcommand{\bracket}[2]{\langle#1|#2\rangle}
\renewcommand{\vec}{\textbf}
\DeclareMathOperator{\id}{\openone}

\begin{document}

\bibliographystyle{apsrev}

\title{Einstein--Podolsky--Rosen correlations of vector bosons}

\author{Pawe{\l}{} Caban}\email{P.Caban@merlin.phys.uni.lodz.pl}
\author{Jakub Rembieli{\'n}ski}\email{jaremb@uni.lodz.pl}
\author{Marta W{\l}odarczyk}\email{marta.wlodarczyk@gmail.com}

\affiliation{Department of Theoretical Physics, University of Lodz\\
Pomorska 149/153, 90-236 {\L}{\'o}d{\'z}, Poland}

\date{\today}

\begin{abstract}
We calculate the joint probabilities and the correlation function in
Einstein--Podolsky--Rosen type experiments with a massive
vector boson in the framework of quantum field theory. We
report on the strange behavior of the correlation
function (and the probabilities) -- the correlation function,
which in the relativistic case still depends on the particle momenta, for
some fixed configurations has local extrema. 
We also show that relativistic spin-1  particles
violate some Bell inequalities more than nonrelativistic ones
and that the degree of violation of the Bell inequality is momentum
dependent. 
\end{abstract}

\pacs{03.65 Ta, 03.65 Ud} 

\maketitle

\section{Introduction}
Different aspects of quantum information theory \cite{nielsen00}
in the relativistic context have been discussed in many papers
\cite{ALH2002, ALHK2003,ALMH2003, AM2002, BT2005, CR2003_Wigner,
CR2005, CR2006,CW2003, Czachor1997_1, Czachor1997_2, Czachor2005,
GA2002, GBA2003, GKM2004, Harshman2005, HSZ2007_1, HSZ2007_2,
JSS2005, JSS2006, KM2003,KS2005, LD2003, LD2004, LMS2005, LPT2003,
LY2004, MAH2003, PS2003, PST2002, PST2005, PT2003_2, PT2003_3,
PT2004_1, RS2002, SL2004, Terno2003, Terno2005, TU2003_1,
TU2003_2, YWYNMX2004, ZBGT2001}, mostly for massive particles.
Photons have only been discussed in a few papers
\cite{AM2002,BT2005,CR2003_Wigner, GBA2003, LPT2003,PT2003_2,
PT2003_3, PT2004_1,Terno2005,TU2003_1}. Most of these works 
were performed in the framework of relativistic quantum mechanics.
However, for the discussion of relativistic covariance the most
appropriate framework is the quantum field theory (QFT) approach. Recently
we have discussed the Einstein--Podolsky--Rosen (EPR) correlation
function for a pair of spin-$\tfrac{1}{2}$ massive particles in
the QFT framework \cite{CR2006}. In the present paper we consider
a pair of spin-1 massive particles in this framework and
calculate the correlation function in EPR-type experiments for
such a pair in a covariant scalar state. We also calculate the
probabilities of the definite outcomes of spin projections
measurements performed by two observers -- Alice and Bob.

We observe very surprising behavior of the correlation function (as
well as the probabilities). In the center-of-mass frame for the definite
configuration of the particles momenta and directions of the spin
projection measurements the correlation function still depends on
the value of the particle momentum. It also appears that for some
configurations this dependence is not monotonic. In other words, for
fixed spin measurement directions and particle momenta directions,
the correlation function (and probabilities) can have an extremum
for some finite value of the particle momentum. As far as we are
aware this is the first time, that such behavior of the correlation
function has been reported.

This strange behavior of the correlation function also
affects the violation of the Bell-type inequalities.
Our analysis shows that relativistic vector bosons
violate Bell inequalities stronger than nonrelativistic spin-1
particles and that the degree of violation of Bell inequality depends
on the particle momentum.

In Sec.~II we establish notation and recall basic facts concerning
the massive spin-1 representation of the Poincar\'e group and
quantum spin-1 boson field. In Sec.~III we define one- and
two-particle states which transform covariantly with respect to
the Lorentz group. In the next section we discuss the spin
operator. Sec.~V is devoted to the explicit calculation of the
probabilities and correlation function for the boson pair in the
scalar state. In Sec.~VI we discuss our correlation
  function and probabilities . 
Sec.~\ref{sec:Bell_inequality} is devoted to the analysis of Bell-type
inequalities for spin 1 particles in the relativistic context.
The last section
contains our concluding remarks.

 In the paper we use the natural units
$\hbar=c=1$ and the metric tensor $\eta^{\mu\nu}=\textrm{diag}(1,-1,-1,-1)$.

\section{Preliminaries}
For the readers convenience we recall the basic facts and formulas
concerning  the spin one representation of the Poincar\'e group
and quantum vector boson field.

\subsection{Massive representations of the Poincar\'e group}
Let us denote by $\mathcal{H}$ the carrier space of the
irreducible massive representation of the Poincar\'e group. It is
spanned by the four-momentum operator eigenvectors
$\ket{k,\sigma}$
\begin{equation}
    \label{eq:dzia³anie_pêdu_na_bazê}
    \hat{P}^{\mu}\ket{k,\sigma}=k^{\mu}\ket{k,\sigma},
\end{equation}
$k^2=m^2$, with $m$ denoting the mass of the particle, and $\sigma$
its spin component along z axis. We use the following
Lorentz-covariant normalization
\begin{equation}
    \label{eq:normalizacja_wektorów_bazowych}
    \bracket{k,\sigma}{k',\sigma'}=2k^0\delta^3(\textbf{k}-\textbf{k}^{\prime})\delta_{\sigma\sigma'}.
\end{equation}
The vectors $\ket{k,\sigma}$ can be generated from standard vector
$\ket{\tilde{k},\sigma}$, where $\tilde{k}=m(1,0,0,0)$ is the
four-momentum of the particle in its rest frame. We have
$\ket{k,\sigma}=U(L_k)\ket{\tilde{k},\sigma}$, where Lorentz boost
$L_k$ is defined by relations $k=L_k\tilde{k}$,
$L_{\tilde{k}}=\id$. The explicit form of $L_k$ is
 \begin{equation}
 \label{eq:Lk jawnie}
 L_k=\left(\begin{array}{c|c}
 \tfrac{k^0}{m} & \tfrac{\bold{k}^T}{m}\\
 \hline
 \tfrac{\bold{k}}{m} & \id+\tfrac{\bold{k}\otimes\bold{k}^T}{m(m+k^0)}
 \end{array}\right),
\end{equation}
where $k^0=\sqrt{m^2+\bold{k}^2}$.

By means of Wigner procedure we get
\begin{equation}
    \label{eq:dzia³anie_U(L)_na_bazê}
    U(\Lambda)\ket{k,\sigma}=\mathcal{D}^s_{\lambda\sigma}(R(\Lambda,k))\ket{\Lambda
    k,\lambda},
\end{equation}
where the Wigner rotation $R(\Lambda,k)$ is defined as
$R(\Lambda,k)=L_{\Lambda k}^{-1}\Lambda L_k$. Because we are going
to analyze correlations of spin one particles, in the sequel we
will focus on the representation
$\mathcal{D}^1(R(\Lambda,k))\equiv\mathcal{D}(R)$. There exists
such unitary matrix $V$ that every matrix $\mathcal{D}(R)$ is
related to $R$ by
\begin{equation}
\label{eq:equivalent}
    \mathcal{D}(R)=VRV^{\dag}.
\end{equation}
$\mathcal{D}(R)$ are generated by $\bold{S}^i$, $i=1,2,3$,
\begin{eqnarray}
    \label{eq:macierze_pauliego_spin_1}
    &&S^1=\tfrac{1}{\sqrt{2}}\left(
                            \begin{array}{ccc}
                              0 & 1 & 0 \\
                              1 & 0 & 1 \\
                              0 & 1 & 0 \\
                            \end{array}
                          \right),\;
    S^2=\tfrac{i}{\sqrt{2}}\left(
                            \begin{array}{ccc}
                              0 & -1 & 0 \\
                              1 & 0 & -1 \\
                              0 & 1 & 0 \\
                            \end{array}
                          \right),\notag\\
   &&\,\,\,\,\, S^3=\left(
          \begin{array}{ccc}
            1 & 0 & 0 \\
            0 & 0 & 0 \\
            0 & 0 & -1 \\
          \end{array}
        \right),
\end{eqnarray}
(see e.g.~\cite{Messiah}), that is
$\mathcal{D}(R)=e^{i\boldsymbol{\phi}\bold{S}}$. Taking into account
the form of generators of the rotations $R$
i.e.~$[I^i]_{jk}=-i\epsilon_{ijk}$, we can easily determine the
explicit form of matrix V
\begin{equation}
    \label{eq:macierzV}
    V=\frac{1}{\sqrt{2}}\left(
      \begin{array}{ccc}
        -1 & i & 0 \\
        0 & 0 & \sqrt{2} \\
        1 & i & 0 \\
      \end{array}
    \right).
\end{equation}

\subsection{Vector field}

Under Lorentz group action the vector boson field operator
$\hat{\varphi}^{\mu}(x)$ transforms according to
\begin{equation}
    \label{eq:transformacja_pola}
    U(\Lambda)\hat{\varphi}^{\mu}(x)U^{\dag}(\Lambda)=(\Lambda^{-1})^{\mu}_{\,\,\nu}\hat{\varphi}^{\nu}(\Lambda
    x).
\end{equation}
The field operator has the standard momentum expansion
\begin{align}
    \label{eq:definicja_pola}
    \hat{\varphi}^{\mu}(x)=(2\pi)^{-3/2}\sum_{\sigma=0,\pm1}\int
    d\mu(k)
    \left[e^{ikx}e^{\mu}_{\,\,\sigma}(k)a_{\sigma}^{\dag}(k)\right.\notag\\
    \left.+e^{-ikx}e^{*\mu}_{\,\,\,\,\,\sigma}(k)a_{\sigma}(k)\right],
\end{align}
where $d\mu(k)=\Theta(k^0)\delta((k^0)^2-\omega_k^2)\equiv
\frac{d^3\textbf{k}}{2\omega_k}$ is the Lorentz-invariant measure,
$\omega_k=\sqrt{\bold{k}^2+m^2}$, $a_{\sigma}^{\dag}(k)$ and
$a_{\sigma}(k)$ are creation and annihilation operators of the
particle with four-momentum $k$ and spin component along z-axis
equal to $\sigma$.
They fulfill canonical commutation relations
\begin{subequations}
\label{eq:reg_komutacyjne_aa+}
    \begin{eqnarray}
        \label{seq:reg_komutacyjne_aa+_zero}
        [a_{\sigma}^{\dag}(k),a_{\sigma'}^{\dag}(k')]&=&[a_{\sigma}(k),a_{\sigma'}(k')]=0,\\
        \label{seq:reg_komutacyjne_aa+_delta}
        [a_{\sigma}(k),a^{\dag}_{\sigma'}(k')]&=&2k^0\delta(\textbf{k}-\textbf{k}')\delta_{\sigma\sigma'}.
    \end{eqnarray}
\end{subequations}
The field satisfies Klein-Gordon equation and Lorentz
transversality condition, which imply
\begin{equation}
    \label{eq:k2=m2_ke}
        m^2=k^2,\quad
        k_{\mu}e^{\mu}_{\,\,\sigma}(k)=0.
\end{equation}
The one-particle states
$\ket{k,\sigma}:=a_{\sigma}^{\dag}(k)\ket{0}$ transform according to
(\ref{eq:dzia³anie_U(L)_na_bazê}) provided that
\begin{subequations}
\label{eq:transformacja_aa+}
    \begin{eqnarray}
    \label{seq:transformacja_a+}
    U(\Lambda)a_{\sigma}^{\dag}(k)U^{\dag}(\Lambda)&=&\mathcal{D}_{\lambda\sigma}(R(\Lambda,k))a_{\lambda}^{\dag}(\Lambda k),\\
    \label{seq:transformacja_a}
    U(\Lambda)a_{\sigma}(k)U^{\dag}(\Lambda)&=&\mathcal{D}^{*}_{\lambda\sigma}(R(\Lambda,k))a_{\lambda}(\Lambda k).
    \end{eqnarray}
\end{subequations}
Here $\ket{0}$ denotes Poincar\'e invariant vacuum with
$\bracket{0}{0}=1$; $a_{\sigma}(k)\ket{0}=0$. Equations
(\ref{eq:transformacja_pola},\ref{eq:transformacja_aa+}) imply the
Weinberg conditions for amplitudes $ e^{\mu}_{\,\,\sigma}(k)$
\begin{equation}
    \label{eq:warunek_Weinberga}
    e^{\mu}_{\,\,\sigma}(\Lambda
    k)=\Lambda^{\mu}_{\,\,\nu}e^{\nu}_{\,\,\lambda}(k)\mathcal{D}
    (R(\Lambda,k))_{\sigma\lambda}.
\end{equation}
From Eq.~(\ref{eq:warunek_Weinberga}) we have
\begin{equation}
    \label{eq:transformacja e(k)}
    e(k)=L_k e(\tilde{k}),
\end{equation}
where $L_k$ is given by (\ref{eq:Lk jawnie}) and we used the fact
that $R(L_k,\tilde{k})=\id$. Therefore, to find the explicit form of
$e^{\mu}_{\,\,\sigma}(k)$ it is enough to determine
$e^{\mu}_{\,\,\sigma}(\tilde{k})$. From Eq.~(\ref{eq:k2=m2_ke}) we
get
\begin{equation}
    \label{seq:emu(ktylda)}
    [e^{\mu}_{\,\,\sigma}(\tilde{k})]=\left(
                                 \begin{array}{ccc}
                                   0&0&0 \\\hline
                                   &\tilde{e}& \\
                                 \end{array}
                               \right),
\end{equation}
where $\tilde{e}$ is a  $3\times 3$ matrix. Now, from the Weinberg
condition (\ref{eq:warunek_Weinberga}) for pure rotations and by
means of Eq.~(\ref{eq:equivalent}) and Schur's Lemma
 we find
\begin{equation}
    \label{eq:e(ktylda_znormalizowane)}
    \tilde{e}=V^{\mathrm{T}},
\end{equation}
where explicit form of $V$ is given by (\ref{eq:macierzV}). Finally
from (\ref{eq:transformacja e(k)}) we have
\begin{equation}
    \label{eq:e(k)}
    e(k)=\left(\begin{array}{c}
 \tfrac{\bold{k}^T}{m}\\
 \hline
 \id+\tfrac{\bold{k}\otimes\bold{k}^T}{m(m+k^0)}
 \end{array}\right)V^{\mathrm{T}}.
\end{equation}
Equations (\ref{eq:e(ktylda_znormalizowane)},\ref{eq:e(k)}) imply
\begin{subequations} \label{seq:relacje_na_amplitudy}
    \begin{equation}
    \label{eq:relacja_e*emu}
        e^{*\mu}_{\,\,\,\,\,\sigma}(k)e_{\mu\lambda}(k)=-\delta_{\sigma\lambda},
    \end{equation}
    \begin{equation}
    \label{eq:relacja_eemu}
        e^{\mu}_{\,\,\sigma}(k)e_{\mu\lambda}(k)=-(VV^{\mathrm{T}})_{\sigma\lambda},
    \end{equation}
    \begin{equation}
    \label{eq:relacja_e*esigma}
       e^{*\mu}_{\,\,\,\,\,\sigma}(k)e^{\nu}_{\,\,\sigma}(k)=-\eta^{\mu\nu}+\tfrac{k^{\mu}k^{\nu}}{m^2},
    \end{equation}
\end{subequations}
where $e(k)VV^{\mathrm{T}}=e^*(k)$, and $VV^{\mathrm{T}}=\left(
                                   \begin{array}{ccc}
                                     0 & 0 & -1 \\
                                     0 & 1 & 0 \\
                                     -1 & 0 & 0 \\
                                   \end{array}
                                 \right)$.

\section{Covariant states}
\subsection{One-particle covariant states}
In the discussion of Lorentz-covariance it is convenient to use
states
\begin{equation}
    \label{eq:stan_kowariantny}
    \ket{(\mu,k)}=e^{\mu}_{\,\,\sigma}(k)\ket{k,\sigma},
\end{equation}
which transform covariantly
\begin{equation}
    \label{eq:stany_kow_transformacje}
    U(\Lambda)\ket{(\mu,k)}=(\Lambda^{-1})^{\mu}_{\,\,\nu}\ket{(\nu,\Lambda
    k)}.
\end{equation}
They are normalized as follows [c.f.~(\ref{eq:normalizacja_wektorów_bazowych})]
\begin{equation}
    \label{eq:stan_kowariantny_iloczyn_skalarny}
    \bracket{(\mu, k)}{(\nu,p)}=2k^0\delta(\bold{k}-\bold{p})e^{*\mu}_{\,\,\,\sigma}(k)e^{\nu}_{\,\,\sigma}(p).
\end{equation}
Arbitrary one-particle state can be expanded in the standard basis
$\ket{k,\sigma}$ as well as in the covariant one
(\ref{eq:stan_kowariantny})
\begin{equation}
    \label{eq:stan_jednoczastkowy_ogólny}
    \ket{\psi}=\int d\mu(k)\psi_{\sigma}(k)\ket{k,\sigma}=\int
    d\mu(k)\Psi_{\mu}(k)\ket{(\mu,k)},
\end{equation}
where
\begin{equation}
    \label{eq:porownanie_amplitud_jedna_czastka}
    \Psi_{\mu}(k)e^{\mu}_{\,\,\sigma}(k)=\psi_{\sigma}(k).
\end{equation}

\subsection{Two-particle covariant states}
In analogy to (\ref{eq:stan_kowariantny}) we can define covariant
basis in the two-particle sector of the Fock space
\begin{equation}
    \label{eq:baza_dwuczastkowa}
    \ket{(\mu,k),(\nu,p)}=e^{\mu}_{\,\,\sigma}(k)e^{\nu}_{\,\,\lambda}(p)\ket{(k,\sigma),(p,\lambda)},
\end{equation}
where $
\ket{(k,\sigma),(p,\lambda)}=a^{\dag}_{\,\sigma}(k)a^{\dag}_{\,\lambda}(p)\ket{0}$.
The most general two-particle state has the form
\begin{multline}
    \label{eq:stan_kowariantny_dwucz¹stkowy}
    \ket{\Psi}=\int
    d\mu(k)d\mu(p)\psi_{\sigma\lambda}(k,p)\ket{(k,\sigma),(p,\lambda)}
    \\\equiv\int
    d\mu(k)d\mu(p)\Psi_{\mu\nu}(k,p)\ket{(\mu,k),(\nu,p)}.
\end{multline}
One can see that
\begin{equation}
    \label{eq:porównanie_amplitud}
    \Psi_{\mu\nu}(k,p)e^{\mu}_{\,\,\sigma}(k)e^{\nu}_{\,\,\lambda}(p)=\psi_{\sigma\lambda}(k,p).
\end{equation}
Moreover it holds
\begin{subequations}
    \label{seq:warunki_na_funkcje_psi}
    \begin{eqnarray}
        \label{eq:warunek_na_psi_sym}
        &&\Psi^{\mu\nu}(k,p)=\Psi^{\nu\mu}(p,k),\\
        \label{eq:warunek_na_psi_poprzecznosc}
        &&k_{\mu}\Psi^{\mu\nu}(k,p)=\Psi^{\mu\nu}(k,p)p_{\nu}=0.
    \end{eqnarray}
\end{subequations}
We can now define two-particle states transforming according to
irreducible representations of Lorentz group. The scalar state
describing particles with sharp momenta is defined as
\begin{equation}
    \label{eq:stan_skalarny}
    \ket{\Psi_s}=\eta_{\mu\nu}\ket{(\mu,k),(\nu,p)}.
\end{equation}
In terms of (\ref{eq:baza_dwuczastkowa}) it takes the form
\begin{equation}
    \label{eq:stan_skalarny_baza1}
    \ket{\Psi_s}
    =\eta_{\mu\nu}e^{\mu}_{\,\,\sigma}(k)e^{\nu}_{\,\,\lambda}(p)\ket{(k\sigma),(p,\lambda)}.
\end{equation}
There are also two independent tensor states, the symmetric
traceless
\begin{equation}
    \label{eq:stan_symetryczny}
    \ket{\Psi_{sym}^{\mu\nu}}=\tfrac{1}{2}(\delta^{\mu}_{\,\,\alpha}\delta^{\nu}_{\,\,\beta}+\delta^{\mu}_{\,\,\beta}\delta^{\nu}_{\,\,\alpha}
    -\tfrac{1}{2}\eta^{\mu\nu}\eta_{\alpha\beta})
    \ket{(\alpha,k),(\beta,p)},
\end{equation}
and the antisymmetric one
\begin{equation}
    \label{eq:stan_antysymetryczny}
    \ket{\Psi_{asym}^{\mu\nu}}=\tfrac{1}{2}(\delta^{\mu}_{\,\,\alpha}\delta^{\nu}_{\,\,\beta}-\delta^{\mu}_{\,\,\beta}\delta^{\nu}_{\,\,\alpha}
    )
    \ket{(\alpha,k),(\beta,p)}.
\end{equation}
In the sequel we will analyze correlations in the scalar state
(\ref{eq:stan_skalarny}).

\section{Spin operator}
When we want to calculate explicitly correlation functions, we need
to introduce the spin operator for relativistic massive particles.
Several possibilities have been discussed in the literature (see
e.g.~\cite{Czachor1997_1,CW2003,Terno2003,RS2002,CR2005, CR2006,
LY2004, LD2003,SL2004, TU2003_1, TU2003_2}). We choose the operator
\begin{equation}
    \label{eq:definicja_operatora_spinu}
    \hat{\bold{S}}=\frac{1}{m}\left(\hat{\bold{W}}+\hat{W}^0\frac{\bold{\hat{P}}}{\hat{P}^0+m}\right),
\end{equation}
which is the most appropriate \cite{Terno2003,CR2005,bogolubov75}.
Here
\begin{equation}
    \label{eq:definicja_wektora_pauliego_lubanskiego}
    \hat{W}^{\mu}=\tfrac{1}{2}\epsilon^{\mu\nu\gamma\delta}\hat{P}_{\nu}\hat{J}_{\gamma\delta}
\end{equation}
is the Pauli-Lubanski four-vector, $\hat{P}_{\nu}$ is the
four-momentum operator, $\hat{J}_{\mu\nu}$ denote the generators of
the Lorentz group such that
$U(\Lambda)=\exp(i\omega^{\mu\nu}\hat{J}_{\mu\nu})$, and we assume
$\epsilon^{0123}=1$. 
Consequently the spin operator $\hat{\bold{S}}$ acts on one-particle
states according to
\begin{equation}
    \label{eq:spin_w_niekowariantnej}
    \hat{\bold{S}}\ket{k,\sigma}=\bold{S}_{\lambda\sigma}\ket{k,\lambda},
\end{equation}
where $S^i$ are defined by (\ref{eq:macierze_pauliego_spin_1}). In
the Fock space $\hat{\bold{S}}$ takes standard form
\begin{equation}
    \label{eq:spin_qft_s}
    \hat{\bold{S}}=\int d\mu(k)a^{\dag}(k)\bold{S}a(k),
\end{equation}
where the column matrix
$a(k)\nolinebreak=\nolinebreak\left(a_{+1}(k),a_0(k),a_{-1}(k)\right)^{\mathrm{T}}$.
From Eqs.~(\ref{eq:relacja_e*emu},\ref{eq:stan_kowariantny}) we get
%
\begin{equation}
    \label{eq:spin_w_bazie_niekow_notacja_macierzowa}
    \bold{\hat{S}}\ket{(\alpha,k)} = 
    - \big[ e(k)\bold{S}^{T} e^{\dag}(k) \eta
    \big]^{\alpha}_{\phantom{\alpha}\beta}
    \ket{(\beta,k)}.
\end{equation}
In real experiments detectors register only particles whose momenta 
belong to some definite region $\Omega$ in momentum space.
Therefore we need the operator which acts similar to
(\ref{eq:spin_w_niekowariantnej}) on particles with four-momenta
belonging to $\Omega$ and yields $0$ in all other cases. Such an
operator has the following form
\begin{equation}
    \label{eq:spin_qft_sa}
    \hat{\bold{S}}_{\Omega}=\int_{\Omega}
    d\mu(k)a^{\dag}(k)\bold{S}a(k).
\end{equation}

\section{Probabilities and the correlation function}
Let us consider two distant observers, Alice and Bob, in the same
inertial  frame, sharing a pair of bosons in scalar state
$\ket{\Psi_s}$ defined by (\ref{eq:stan_skalarny_baza1}).

Now let Alice measure spin component of her boson in direction
$\bold{a}$ and Bob spin component of his boson in direction
$\bold{b}$, where $|\bold{a}|=|\bold{b}|=1$. Their observables are
$(\bold{a}\cdot\bold{\hat{S}}_A)$ and $(\bold{b}\cdot\bold{\hat{S}}_B)$,
respectively, where $(\boldsymbol{\omega}\cdot\hat{\bold{S}}_{\Omega})$ is
defined by (\ref{eq:spin_qft_sa}) with $\Omega$ equal $A$ and $B$,
and $\boldsymbol{\omega}$ equal to $\vec{a}$ and $\vec{b}$,
respectively. We assume that $A\cap B=\emptyset$. Now we would like
to explicitly calculate probabilities $P_{\sigma\lambda}$ of
obtaining particular outcomes $\sigma$ and $\lambda$ by Alice and
Bob, respectively ($\sigma$ and $\lambda$ can take values $\pm1$ and
$0$). Let us first notice, that from
Eqs.~(\ref{eq:baza_dwuczastkowa}),
(\ref{eq:spin_w_niekowariantnej}), and  (\ref{eq:spin_qft_sa}) we
have
\begin{multline}
\label{eq:orozkladzie_dzialanie_omegasomega}
    (\boldsymbol{\omega}\cdot\hat{\bold{S}}_{\Omega})
    \ket{(k,\lambda),(p,\sigma)}\\
   = \boldsymbol{\omega} \cdot \big[ \chi_{\Omega}(k) 
     \bold{S}_{\lambda'\lambda} \delta_{\sigma'\sigma} +
    \chi_{\Omega}(p) \bold{S}_{\sigma'\sigma} \delta_{\lambda'\lambda}
    \big] \ket{(k,\lambda'),(p,\sigma')},
\end{multline}
where the characteristic function $\chi_{\Omega}(q)$ is defined in a
standard way
\begin{equation}
    \label{eq:funcja_charakterystyczna}
    \chi_{\Omega}(q)=\left\{
                                 \begin{array}{ccccc}
                                   1 &  &  &  & \text{when}\,\, q\in\Omega,\\
                                   0 &  &  &  & \text{when}\,\,q \notin\Omega.\\
                                 \end{array}
                               \right.
\end{equation}
However, in EPR-type experiments we take into account only such
measurements in which Alice and Bob register one particle each.
Therefore we are actually interested in spectral decomposition of
observable
$(\boldsymbol{\omega}\cdot\hat{\bold{S}}_{\Omega}) \hat{\Pi}^{1}_{\Omega}$,
where $\hat{\Pi}^{1}_{\Omega}$ is a projector (in the two-particle
sector of the Fock space) on the subspace of states corresponding to
the situation in which exactly one particle has momentum from the
region $\Omega$. To find the explicit form of the
$\hat{\Pi}^{1}_{\Omega}$ we use the particle number operator
$\hat{N}_{\Omega}$ answering the question how many particles have
momentum from $\Omega$. In the two-particle sector of the Fock space
we have the obvious spectral decomposition of $\hat{N}_{\Omega}$
\begin{equation}
\label{eq:orozkladzie_rozklad_operatora_liczby_czastek}
    \hat{N}_{\Omega} = 0\cdot\hat{\Pi}^0_{\Omega}+1\cdot\hat{\Pi}^1_{\Omega}+2\cdot\hat{\Pi}^2_{\Omega},
\end{equation}
and in the basis (\ref{eq:baza_dwuczastkowa})
\begin{equation}
\label{eq:orozkladzie_operator_lczastek_baza}
    \hat{N}_{\Omega}\ket{(k,\lambda),(p,\sigma)} = 
    \big[\chi_{\Omega}(k)+\chi_{\Omega}(p)\big]
    \ket{(k,\lambda),(p,\sigma)}.
\end{equation}
In Eq.~(\ref{eq:orozkladzie_rozklad_operatora_liczby_czastek})
$\hat{\Pi}_{\Omega}^i$, $i=0,1,2$, denotes a projector on the
subspace of two-particle states, in which exactly $i$ particles have
momenta from $\Omega$. From
Eqs.~(\ref{eq:orozkladzie_rozklad_operatora_liczby_czastek},
\ref{eq:orozkladzie_operator_lczastek_baza}) we find
\begin{equation}
    \label{eq:orozkladzie_piomegajeden}
    \hat{\Pi}_{\Omega}^1 = 2\hat{N}_{\Omega}-\hat{N}_{\Omega}^2,
\end{equation}
and
\begin{multline}
    \label{eq:orozkladzie_piomegajeden_wbazie}
    \hat{\Pi}_{\Omega}^1 \ket{(k,\lambda),(p,\sigma)} \\
    = \big[ \chi_{\Omega}(k) + \chi_{\Omega}(p) - 2 \chi_{\Omega}(k) 
      \chi_{\Omega}(p)\big] \ket{(k,\lambda),(p,\sigma)}.
\end{multline}
Therefore from (\ref{eq:orozkladzie_dzialanie_omegasomega}) and
(\ref{eq:orozkladzie_piomegajeden_wbazie}) we finally get
\begin{multline}
    \label{eq:orozkladzie_dzialanie_omegasomega_piomegajeden}
    (\boldsymbol{\omega}\cdot\hat{\bold{S}}_{\Omega}) \Pi^{1}_{\Omega}
    \ket{(k,\lambda),(p,\sigma)} \\
    = \boldsymbol{\omega} \cdot \Big\{
    \chi_{\Omega}(k)\big[1-\chi_{\Omega}(p)\big]
    \bold{S}_{\lambda'\lambda} \delta_{\sigma'\sigma} \\
    + \chi_{\Omega}(p) \big[1-\chi_{\Omega}(k)\big]
    \bold{S}_{\sigma'\sigma} \delta_{\lambda'\lambda} \Big\}
    \ket{(k,\lambda'),(p,\sigma')}.
\end{multline}
By definition the observable
$\boldsymbol{\omega}\hat{\bold{S}}_{\Omega}\hat{\Pi}^{1}_{\Omega}$
measures the spin component of one particle in the direction
$\boldsymbol{\omega}$, therefore its spectral decomposition is
\begin{equation}
    \label{eq:orozkladzie_rozklad_omegasomegapiomegajeden}
    (\boldsymbol{\omega}\cdot\hat{\bold{S}}_{\Omega})\hat{\Pi}^{1}_{\Omega}
    =1 \cdot \hat{\Pi}^{+}_{\Omega\boldsymbol{\omega}}
    -1 \cdot \hat{\Pi}^{-}_{\Omega\boldsymbol{\omega}} + 
     0 \cdot \hat{\Pi}^{0}_{\Omega\boldsymbol{\omega}},
\end{equation}
where the projectors $\hat{\Pi}^{\pm}_{\Omega\boldsymbol{\omega}}$
and $\hat{\Pi}^{0}_{\Omega\boldsymbol{\omega}}$ correspond to
eigenvalues $\pm1$ and $0$, respectively. Simple calculation gives
\begin{subequations}
\label{seq:orozkladzie_projektory}
    \begin{eqnarray}
        \label{eq:orozkladzie_piplusminus}
        \hat{\Pi}^{\pm}_{\Omega\boldsymbol{\omega}} & = & 
        \tfrac{1}{2}
        (\boldsymbol{\omega}\cdot\hat{\bold{S}}_{\Omega})
        \big[(\boldsymbol{\omega}\cdot\hat{\bold{S}}_{\Omega}) \pm \id\big]
        \hat{\Pi}_{\Omega}^1,\\
        \label{eq:orozkladzie_pizero}
        \hat{\Pi}^{0}_{\Omega\boldsymbol{\omega}} & = & 
        \big[\id-(\boldsymbol{\omega}\cdot\hat{\bold{S}}_{\Omega})^2\big]
        \hat{\Pi}_{\Omega}^1.
    \end{eqnarray}
\end{subequations}

Now we can find explicitly the probabilities $P_{\sigma\lambda}$
mentioned above in the state (\ref{eq:stan_skalarny_baza1}).
\begin{eqnarray}
    \label{eq:prawdopodobienstwa_wzory_qm}
    P_{\sigma\lambda}&=&\frac{\bra{\Psi_s}\hat{\Pi}_{A\bold{a}}^{\sigma}\hat{\Pi}_{B\bold{b}}^{\lambda}
    \ket{\Psi_s}}{\bracket{\Psi_s}{\Psi_s}}.
\end{eqnarray}
From Eqs.~(\ref{eq:orozkladzie_dzialanie_omegasomega},
\ref{eq:orozkladzie_piomegajeden_wbazie},
\ref{eq:orozkladzie_dzialanie_omegasomega_piomegajeden},
\ref{seq:orozkladzie_projektory}) we find
\begin{widetext}
\begin{subequations}
    \label{seq:orozkladzie_dzialanie_pi+-pi0}
    \begin{multline}
        \label{eq:orozkladzie_dzialaniepi+-}
        \hat{\Pi}^{\pm}_{\Omega\boldsymbol{\omega}}\ket{\Psi_s}=
        \tfrac{1}{2} \eta_{\mu\nu} 
          e^{\mu}_{\phantom{\mu}\lambda}(k) e^{\nu}_{\phantom{\nu}\sigma}(p)
             \Big\{\big[(\boldsymbol{\omega}\cdot\vec{S})^2
        \pm\boldsymbol{\omega}\cdot\vec{S}\big]_{\lambda'\lambda}
           \delta_{\sigma'\sigma} \chi_{\Omega}(k) 
           \big[1-\chi_{\Omega}(p)\big]\\
        + \big[(\boldsymbol{\omega}\cdot\vec{S})^2
        \pm\boldsymbol{\omega}\cdot\vec{S}\big]_{\delta'\delta}
        \delta_{\lambda'\lambda} \chi_{\Omega}(p)\big[1-\chi_{\Omega}(k)\big]
        \Big\} \ket{(k,\lambda^{\prime}),(p,\sigma')},
    \end{multline}
    \begin{multline}
        \label{eq:orokzladzie_dzialaniepi0}
        \hat{\Pi}^{0}_{\Omega\boldsymbol{\omega}}\ket{\Psi_s} =
        \eta_{\mu\nu}
        e^{\mu}_{\phantom{\mu}\lambda}(k) e^{\nu}_{\phantom{\nu}\sigma}(p)
        \Big\{ \delta_{\lambda'\lambda} \delta_{\sigma'\sigma}
        \big[\chi_{\Omega}(k)+\chi_{\Omega}(p)\big]^2 - 
             (\boldsymbol{\omega}\cdot\vec{S})^2_{\lambda'\lambda}
        \delta_{\sigma'\sigma} \chi_{\Omega}(k)
        \big[1-\chi_{\Omega}(p)\big] \\
        - (\boldsymbol{\omega}\cdot\vec{S})^2_{\sigma'\sigma}
          \delta_{\lambda'\lambda} \chi_{\Omega}(p)
        \big[1-\chi_{\Omega}(k)\big] \Big\} 
        \ket{(k,\lambda'),(p,\sigma')}.
    \end{multline}
\end{subequations}
\end{widetext}
Let us assume, that Alice can measure only the bosons with
four-momentum $k$ and Bob those with four-momentum $p$, i.e.
\begin{equation}
    \label{eq:fajny_warunek_boli_mnie_g³owa}
    \chi_A(p)=\chi_B(k)=0,
\end{equation}
and
\begin{equation}
    \label{eq:fajny_warunek__dzis_nie_boli}
    \chi_A(k)=\chi_B(p)=1.
\end{equation}
After a little algebra we find
\begin{subequations}
\label{eq:prawdopodob_skrocona_forma}
\begin{eqnarray}
    P_{\pm\pm} & = & \frac{1}{4\big[2+\tfrac{(kp)^2}{m^4}\big]}
      \textrm{Tr} \{M(\bold{k,a})\eta M(\bold{p,b})\eta \notag\\
      & & - N(\bold{k,a})\eta N(\bold{p,b})\eta\},\\
    P_{\pm\mp} & = & \frac{1}{4\big[2+\tfrac{(kp)^2}{m^4}\big]}
      \textrm{Tr} \{M(\bold{k,a})\eta M(\bold{p,b})\eta \notag\\
      & & + N(\bold{k,a})\eta N(\bold{p,b})\eta\},\\
    P_{0\pm} & = & \frac{1}{2\big[2+\tfrac{(kp)^2}{m^4}\big]}
    \textrm{Tr} \{T(\bold{k,a})\eta M(\bold{p,b})\eta\},\\
    P_{\pm 0} & = & \frac{1}{2\big[2+\tfrac{(kp)^2}{m^4}\big]}
    \textrm{Tr} \{M(\bold{k,a})\eta T(\bold{p,b})\eta\},\\
    P_{00} & = & \frac{1}{2+\tfrac{(kp)^2}{m^4}}
    \textrm{Tr} \{T(\bold{k,a})\eta T(\bold{p,b})\eta\},
\end{eqnarray}
\end{subequations}
where  we have introduced the following notation
\begin{subequations}
    \label{eq:nowe_macierze_nmt}
    \begin{eqnarray}
    N(\bold{q},\boldsymbol{\omega})^{\alpha\beta} & \equiv & 
          e^{*\alpha}_{\phantom{*\alpha}\lambda}(q)
    (\boldsymbol{\omega}\cdot\vec{S})_{\lambda\sigma}
          e^{\beta}_{\phantom{\beta}\sigma}(q),\\
    M(\bold{q},\boldsymbol{\omega})^{\alpha\beta} & \equiv & 
          e^{*\alpha}_{\phantom{*\alpha}\lambda}(q)
    (\boldsymbol{\omega}\cdot\vec{S})^2_{\lambda\sigma}
          e^{\beta}_{\phantom{\beta}\sigma}(q),\\
    T(\bold{q},\boldsymbol{\omega})^{\alpha\beta} & \equiv & 
          e^{*\alpha}_{\phantom{*\alpha}\lambda}(q)
    \big[\delta_{\lambda\sigma} - 
    (\boldsymbol{\omega}\cdot\vec{S})^2_{\lambda\sigma}\big]
      e^{\beta}_{\phantom{\beta}\sigma}(q),
    \end{eqnarray}
\end{subequations}
(for explicit form of matrices $N$, $M$, $T$ see App.~A.). All the
above probabilities add up to $1$.

Using Eqs.~(\ref{eq:prawdopodob_skrocona_forma}) and
({\ref{seq:nowe_macierze_jawnie}}) one can easily find explicit form
of the probabilities for arbitrary $\bold{a}$, $\bold{b}$,
$\bold{k}$ and $\bold{p}$. However, the resulting formulas appear to
be rather long and we do not put them here. In the next section we
are going to limit ourselves to the simpler case when Alice and Bob
are at rest with respect to the center of mass frame (CMF) of the
boson pair.

In EPR-type experiments we usually analyze the spin correlation
function defined as
\begin{equation}
    \label{eq:f_korlacji_def}
    C_{\bold{ab}}=\sum_{\sigma,\lambda=-s}^{s}\lambda\sigma P_{\lambda\sigma},
\end{equation}
where $\sigma$, $\lambda$ denote spin projections on the directions
$\bold{a}$ and $\bold{b}$, respectively, and $P_{\lambda\sigma}$ is
the joint probability of obtaining results $\sigma$, $\lambda$. Let
us notice that cases when $\sigma$ or $\lambda$ equal $0$ do not
contribute to the correlation function (\ref{eq:f_korlacji_def}). In
principle one could define the ''normalized'' correlation function
as $C_{\bold{ab}}$ [Eq.~(\ref{eq:f_korlacji_def})] divided by
$\sum_{\sigma,\lambda\neq 0}P_{\lambda\sigma}$. However we prefer to
deal with the function (\ref{eq:f_korlacji_def}) which contains more
information. The ''normalized'' correlation function can be also
easily calculated by means of
Eqs.~(\ref{eq:prawdopodob_skrocona_forma}). Therefore in our case
($s=1$) the correlation function takes the following form
\begin{equation}
    \label{eq:f_korlacji_def_z_prawdopodobienstw}
    C_{\bold{ab}}(\bold{k,p}) = P_{++} + P_{--} - P_{-+} - P_{+-},
\end{equation}
which in notation (\ref{eq:nowe_macierze_nmt}) reads
\begin{equation}
    \label{eq:f_korlacji_nowanotacja}
    C_{\bold{ab}}(\bold{k,p})=-\frac{1}{2+\tfrac{(kp)^2}{m^4}}
    \textrm{Tr} \{N(\bold{k,a})\eta N(\bold{p,b})\eta\}.
\end{equation}
Of course the above correlation function can be also found by means
of standard formula
\begin{equation}
    \label{eq:funkcja_korelacji_definicja}
    C_{\bold{ab}}(\bold{k,p})=\frac{\bra{\Psi_s}(\vec{a}\cdot\bold{\hat{S}}_A)
  (\vec{b}\cdot\bold{\hat{S}}_B)\ket{\Psi_s}}
    {\bracket{\Psi_s}{\Psi_s}}.
\end{equation}
After some calculation we get
\begin{multline}
    \label{eq:f_korelacji_jawna}
    C_{\bold{ab}}(\bold{k,p}) \\ = 
       \frac{2}{2+\tfrac{(kp)^2}{m^4}} \Big\{ - \vec{a}\cdot\vec{b} 
    - \frac{[\vec{a}\cdot(\vec{k}\times\vec{p})]
      [\vec{b}\cdot(\vec{k}\times\vec{p})]}{m^2(m+k^0)(m+p^0)} \\
    - \frac{(\vec{a}\cdot\vec{p})(\vec{b}\cdot\vec{k}) 
          - (\vec{a}\cdot\vec{b})(\vec{p}\cdot\vec{k})}{m^2} \\
    +\frac{(\vec{a}\cdot\vec{p})(\vec{b}\cdot\vec{p}) - 
           \vec{p}^2(\vec{a}\cdot\vec{b})}{m(m+p^0)} 
  + \frac{(\vec{a}\cdot\vec{k})(\vec{b}\cdot\vec{k}) - 
             \vec{k}^2(\vec{a}\cdot\vec{b})}{m(m+k^0)} \\
    + \frac{(\vec{k}\cdot\vec{p})(\vec{a}\cdot\vec{p})
                       (\vec{b}\cdot\vec{k})-
    (\vec{k}\cdot\vec{p})^2 (\vec{a}\cdot\vec{b})}{m^2(m+k^0)(m+p^0)}\Big\}.
\end{multline}
In the next section we will analyze behavior of the probabilities
and the correlation function in the CMF frame.

\section{Probabilities and correlation function in CMF frame}
\label{sekcjaV}  In the CMF frame $\vec{p}=-\vec{k}$ and
probabilities (\ref{eq:prawdopodob_skrocona_forma}) take the form
\begin{subequations}
\label{seq:prawdopodob_jawnie_CMF}
\begin{eqnarray}
    P_{\pm\pm} & = & \tfrac{1}{4[2+(1+2x)^2]}
    \Big\{(1+2x)^2-2(1+2x)(\vec{a}\cdot\vec{b})\notag\\
        & & + 4x (\vec{a}\cdot\vec{n})(\vec{b}\cdot\vec{n})
    -4x(x+1)\big[ (\vec{a}\cdot\vec{n})^2 + 
               (\vec{b}\cdot\vec{n})^2 \big]\notag\\
        & & + \big[ \vec{a}\cdot\vec{b} + 
        2x(\vec{a}\cdot\vec{n}) (\vec{b}\cdot\vec{n}) \big]^2 \Big\},\\
    P_{\pm\mp} & = & \tfrac{1}{4[2+(1+2x)^2]}
          \Big\{(1+2x)^2+2(1+2x)(\vec{a}\cdot\vec{b})\notag\\
                 &&-4x(\vec{a}\cdot\vec{n})(\vec{b}\cdot\vec{n})
    -4x(x+1)\big[(\vec{a}\cdot\vec{n})^2+(\vec{b}\cdot\vec{n})^2\big]\notag\\
               &&+\big[\vec{a}\cdot\vec{b} + 
           2x(\vec{a}\cdot\vec{n})(\vec{b}\cdot\vec{n})\big]^2\Big\},\\
    P_{0\pm} & = & \tfrac{1}{2[2+(1+2x)^2]}
    \Big\{1+4x(1+x)(\vec{a}\cdot\vec{n})^2\notag\\
       & & -\big[\vec{a}\cdot\vec{b} + 
            2x(\vec{a}\cdot\vec{n})(\vec{b}\cdot\vec{n})\big]^2\Big\},\\
    P_{\pm0} & = & \tfrac{1}{2[2+(1+2x)^2]}
    \Big\{1+4x(1+x)(\vec{b}\cdot\vec{n})^2\notag\\
       & & -\big[\vec{a}\cdot\vec{b} + 2x(\vec{a}\cdot\vec{n}) 
             (\vec{b}\cdot\vec{n})\big]^2\Big\},\\
    P_{00} & = & \tfrac{1}{2+(1+2x)^2}
    \big[\vec{a}\cdot\vec{b}+
          2x(\vec{a}\cdot\vec{n})(\vec{b}\cdot\vec{n})\big]^2,
\end{eqnarray}
\end{subequations}
where $x=\Big(\tfrac{|\bold{k}|}{m}\Big)^2$,
$\bold{n}=\tfrac{\bold{k}}{|\bold{k}|}$. Furthermore in this frame
the correlation function reduces to
\begin{multline}
\label{eq:f_korelacji_CMF}
    C_{\bold{ab}}(\bold{k},-\bold{k})=\\\frac{2}{2+(1+2x)^2}
    \big[-(1+2x)(\vec{a}\cdot\vec{b}) 
    +2x(\vec{a}\cdot\vec{n})(\vec{b}\cdot\vec{n})\big]
\end{multline}
For a given configuration of directions $\vec{a}$, $\vec{b}$ and
$\vec{n}$ the probabilities and the correlation function depend on
the value of the three-momentum of the particles. What is very
unexpected, for some configurations the probabilities and the
correlation function have local extrema. 
It suggests that for some values
of momenta Bell inequalities may be violated stronger.
We discuss this possibility in the next section.

\begin{figure}
\centering
\includegraphics[scale=0.7]{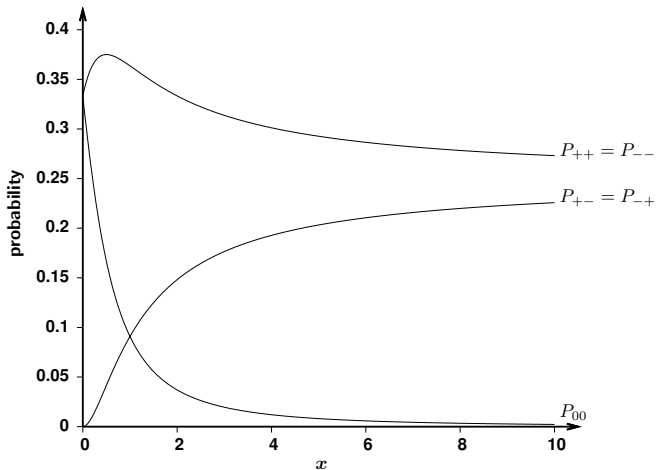}
\caption{The plot shows dependence of probabilities
$P_{\sigma\lambda}$ in CMF frame on $x$ for $\vec{a}\cdot\vec{b}=-1$,
$\vec{a}\cdot\vec{n}=\vec{b}\cdot\vec{n}=0$. 
The probabilities $P_{++}$ and $P_{--}$
have maximum equal to $3/8$ for $x=1/2$. Probabilities $P_{0\pm}$
and $P_{\pm 0}$ vanish.} \label{fig:-100}
\end{figure}

\begin{figure}
\centering
\includegraphics[scale=0.70]{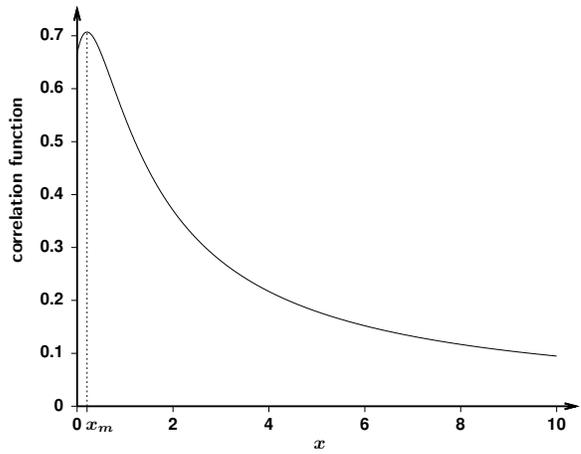}
\caption{The plot shows dependence of correlation function
$C_{\bold{a}\bold{b}}$ in CMF frame  on $x$ for $\vec{a}\cdot\vec{b}=-1$,
$\vec{a}\cdot\vec{n}=\vec{b}\cdot\vec{n}=0$. The function has maximum equal to
$1/\sqrt{2}$ for $x_m=(\sqrt{2}-1)/2$. } \label{fig:fun-100}
\end{figure}

\begin{figure}
\centering
\includegraphics[scale=0.70]{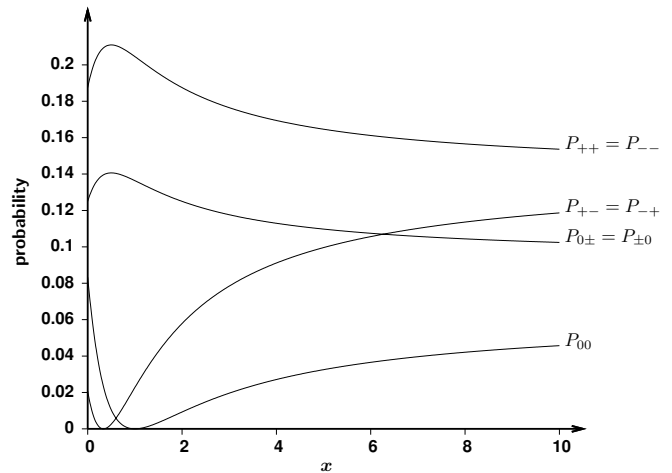}
\caption{The plot shows dependence of probabilities
$P_{\sigma\lambda}$ in CMF frame on $x$ for $\vec{a}\cdot\vec{b}=-1/2$,
$\vec{a}\cdot\vec{n}=\vec{b}\cdot\vec{n}=1/2$. 
The probabilities $P_{++}$, $P_{--}$,
$P_{0\pm}$ and $P_{\pm 0}$ have maxima for $x=1/2$ while
probabilities $P_{00}$, $P_{-+}$ and $P_{--}$ have minima for $x=1$
and $x=1/3$, respectively.} \label{fig:1/2}

\end{figure}

\begin{figure}
\centering
\includegraphics[scale=0.70]{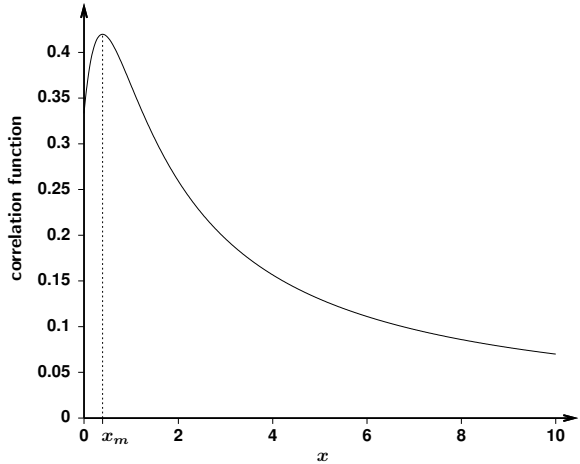}
\caption{The plot shows dependence of the correlation function
$C_{\bold{a}\bold{b}}$ in CMF frame  on $x$ for $\vec{a}\cdot\vec{b}=-1/2$,
$\vec{a}\cdot\vec{n}=\vec{b}\cdot\vec{n}=1/2$. 
It has maximum equal to $(\sqrt{19}-1)/8$
for $x_m=(\sqrt{19}-2)/6$.} \label{fig:fun_1/2}
\end{figure}

Configurations can be found where the correlation
function and some of the probabilities have local extrema, while
other probabilities are monotonic (see Figs.~\ref{fig:-100},
\ref{fig:fun-100}).
Configurations can also be found, where all the
probabilities are monotonic and such configurations where all of the
probabilities and the correlation function have local extrema (see
Figs.~\ref{fig:1/2}, \ref{fig:fun_1/2}).


Finally let us consider the ultra-relativistic
($x\longrightarrow\infty$) and non-relativistic ($x\longrightarrow
0$) limits of formulas (\ref{seq:prawdopodob_jawnie_CMF}) and
(\ref{eq:f_korelacji_CMF}).

\subsubsection*{Ultra-relativistic limit}

In the ultra-relativistic limit the probabilities take the form
\begin{subequations}
\label{seq:prawdopodob_ultrarel}
\begin{eqnarray}
        P_{\pm\pm}=P_{\pm\mp} & = & \tfrac{1}{4} 
             (1-(\vec{a}\cdot\vec{n})^2)(1-(\vec{b}\cdot\vec{n})^2),\\
        P_{0\pm} & = & 
             \tfrac{(\vec{a}\cdot\vec{n})^2}{2}(1-(\vec{b}\cdot\vec{n})^2),\\
        P_{\pm0}&=&\tfrac{(\vec{b}\cdot\vec{n})^2}{2}(1-(\vec{a}\cdot\vec{n})^2),\\
        P_{00}&=&(\vec{a}\cdot\vec{n})^2(\vec{b}\cdot\vec{n})^2,
    \end{eqnarray}
\end{subequations}
and the correlation function vanishes
\begin{equation}
    \label{eq:f_korelacji_ultrarel}
    C_{\bold{ab}}(\vec{k},-\vec{k})=0,
\end{equation}
which means that for ultra fast particles there is no correlation
between outcomes of measurements performed by Alice and Bob. One can
notice that in this limit none of the probabilities
(\ref{seq:prawdopodob_ultrarel}) depend on relative configuration of
directions $\vec{a}$ and $\vec{b}$ but only on their configuration
with respect to direction of the momentum $\vec{n}$.
Fig.~\ref{fig:array} illustrates the dependence of probabilities
(\ref{seq:prawdopodob_ultrarel}) on scalar products $\vec{a}\cdot\vec{n}$ and
$\vec{b}\cdot\vec{n}$.
\begin{widetext}
\begin{center}
\begin{figure}
\includegraphics[scale=1.25]{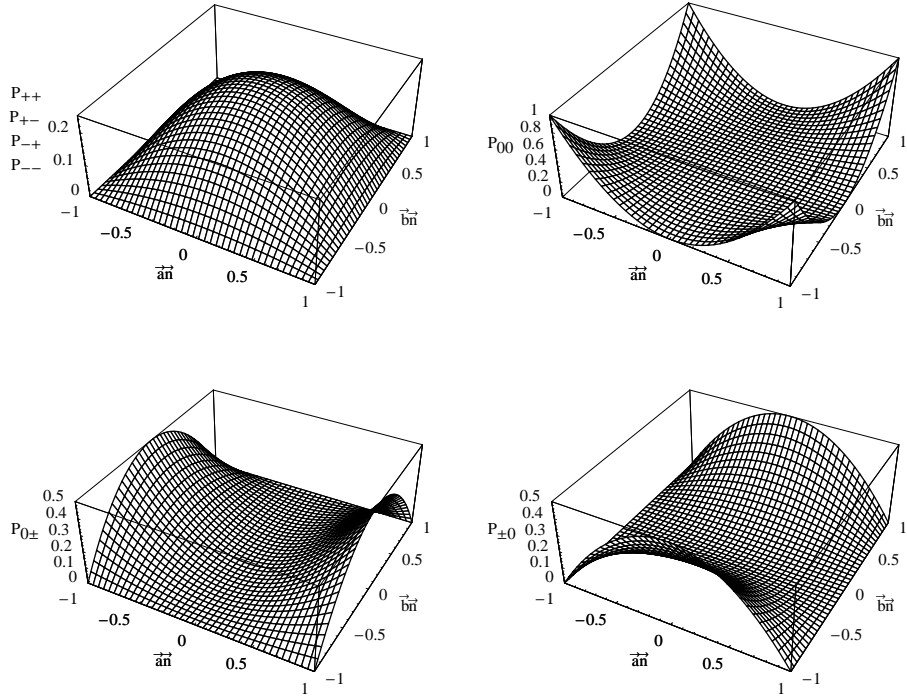}
\caption{Dependence of probabilities
(\ref{seq:prawdopodob_ultrarel}) on $\vec{a}\cdot\vec{n}$ and
$\vec{b}\cdot\vec{n}$ in 
the ultra-relativistic limit.} \label{fig:array}
\end{figure}
\end{center}
\end{widetext}

\subsubsection*{Non-relativistic limit}

Now in non-relativistic limit the probabilities are
\begin{subequations}
\label{seq:prawdopodob_nierel_granica}
    \begin{eqnarray}
        P_{\pm\pm}&=&\tfrac{1}{12}(1-(\vec{a}\cdot\vec{b}))^2,\\
        P_{\pm\mp}&=&\tfrac{1}{12}(1+(\vec{a}\cdot\vec{b}))^2,\\
        P_{0\pm}=P_{\pm 0}&=&\tfrac{1}{6}(1-(\vec{a}\cdot\vec{b})^2),\\
        P_{00}&=&\tfrac{1}{3}(\vec{a}\cdot\vec{b})^2,
    \end{eqnarray}
\end{subequations}
and the correlation function reads
\begin{equation}
    \label{eq:f_korelacji_jawna_nierel}
    C_{\bold{ab}}=-\tfrac{2}{3}\vec{a}\cdot\vec{b}.
\end{equation}
Let us note that in this limit probabilities and the correlation
function do not depend on the momentum $\bold{k}$. One can also
easily check that in this case they are the same as calculated in
the framework of non-relativistic quantum mechanics in the singlet
state
\begin{equation}
    \label{eq:stan_skalarny_nierel}
    \ket{\Psi}=
    \tfrac{1}{\sqrt{3}}(\ket{1}\ket{-1}-\ket{0}\ket{0}+\ket{-1}\ket{1}), 
\end{equation}
where $\ket{1}$, $\ket{0}$ and  $\ket{-1}$ are states with spin
component along z-axis equal to $1$, $0$ and $-1$, respectively.

\section{Bell-type inequalities}
\label{sec:Bell_inequality}

The spin-1 system has three degrees of freedom, which
makes the full analysis of Bell inequalities much more difficult and
subtle (see e.g.~\cite{zukowski_01, zukowski_01a, HHHH}).
In the present paper we will show that at least 
for some Bell-type inequalities its violation strongly depends on the
particle momenta. Moreover we discuss
inequality which is satisfied for the nonrelativistic correlation
function but is violated in the relativistic case.

For spin-$\tfrac{1}{2}$ particles the most commonly discussed
Bell-type inequality is the Clauser-Horne-Shimony-Holt (CHSH)
inequality \cite{cab_CHSH1969}: 
 \begin{equation}
 |C_{\bold{ab}}-C_{\bold{ad}}|+|C_{\bold{cb}}+C_{\bold{cd}}| \le 2
 \label{eq:CHSH} 
 \end{equation}
In (\ref{eq:CHSH}) $C_{\bold{ab}}$ denotes the
correlation function of spin projections on the directions $\vec{a}$
and $\vec{b}$. One can easily check that (\ref{eq:CHSH}) is
also valid for spin-1 particles.
(see e.g.\ \cite{cab_Ballentine1998}).
The nonrelativistic correlation function
(\ref{eq:f_korelacji_jawna_nierel}) does not violate the inequality
(\ref{eq:CHSH}). Indeed, inserting (\ref{eq:f_korelacji_jawna_nierel})
into (\ref{eq:CHSH}) we get
 \begin{equation}
 |\vec{a}\cdot\vec{b} - \vec{a}\cdot\vec{d}| + |\vec{c}\cdot\vec{b} +
 \vec{c}\cdot\vec{d}| \le 3.
 \label{eq:CHSH_nirel_spin1}
 \end{equation}
The largest value of the left side of (\ref{eq:CHSH_nirel_spin1}) is
equal to $2\sqrt{2}$, therefore (\ref{eq:CHSH_nirel_spin1}) holds in
all configurations. In the relativistic framework, inserting
(\ref{eq:f_korelacji_CMF}) into (\ref{eq:CHSH}) we get the following
inequality:
 \begin{multline}
 \frac{1}{2+(1+2x)^2}\Big\{ \Big| 
 (1+2x)\big[(\vec{a}\cdot\vec{b}) - (\vec{a}\cdot\vec{d})\big]\\
 - 2x\big[(\vec{a}\cdot\vec{n})(\vec{b}\cdot\vec{n}) - 
  (\vec{a}\cdot\vec{n})(\vec{d}\cdot\vec{n})\big] \Big| \\ +
 \Big| (1+2x)\big[(\vec{c}\cdot\vec{b}) + (\vec{c}\cdot\vec{d})\big]\\
 - 2x\big[(\vec{c}\cdot\vec{n})(\vec{b}\cdot\vec{n}) + 
  (\vec{c}\cdot\vec{n})(\vec{d}\cdot\vec{n})\big] 
 \Big| \Big\} \le 1.
 \label{eq:CHSH_ralat_spin1}
 \end{multline}
Our numerical simulations show that the largest value of left side of
(\ref{eq:CHSH_ralat_spin1}) is equal to 1. Therefore
the CHSH inequality is not violated in the relativistic
framework, either. 

Therefore for spin 1 particles we have to consider other Bell-type
inequalities. 

According to Mermin's paper \cite{cab_Mermin1980},
in EPR-type experiments with a pair of spin-1 particles in the
singlet state
the following inequality has to be satisfied 
 \begin{equation}
 C_{\bold{ab}} + C_{\bold{bc}} + C_{\bold{ca}} \le1,
 \label{eq:Bell_inequality_1}
 \end{equation}
in the theory which fulfills the assumptions of local realism.
This inequality, similar to the CHSH one, is not violated in the
nonrelativistic quantum mechanics. Indeed,
inserting (\ref{eq:f_korelacji_jawna_nierel}) into
(\ref{eq:Bell_inequality_1}) we get the inequality
 \begin{equation}
 -\frac{2}{3}(\vec{a}\cdot\vec{b} + \vec{b}\cdot\vec{c} +
 \vec{c}\cdot\vec{a}) \le 1
 \label{eq:Bell_inequality_2}
 \end{equation}
which is equivalent to
 \begin{equation}
 \frac{1}{3}\big[3 - (\vec{a} + \vec{b} + \vec{c})^2\big] \le 1.
 \label{eq:Bell_inequality_3}
 \end{equation}
The left side of (\ref{eq:Bell_inequality_3}) is largest when
$\vec{a}+\vec{b}+\vec{c}=\vec{0}$. In this case (\ref{eq:Bell_inequality_3})
is of course fulfilled. Therefore nonrelativistic quantum mechanics
does not violate the Bell-type inequality
(\ref{eq:Bell_inequality_1}). 

\begin{figure}
\centering
\includegraphics[scale=0.70]{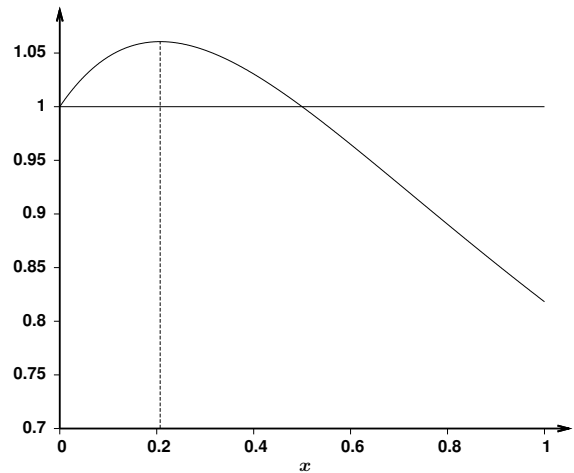}
\caption{The plot shows dependence of the 
left side of the inequality (\ref{eq:Bell_inequality_5}) on $x$.
The plotted function has maximum value equal to $3\sqrt{2}/4$
for $x=(\sqrt{2}-1)/2$.} \label{fig:Bell_1}
\end{figure}

However, we show that the inequality (\ref{eq:Bell_inequality_1}) can
be violated in the relativistic framework. Inserting 
(\ref{eq:f_korelacji_CMF}) into inequality
(\ref{eq:Bell_inequality_1}) we get
 \begin{multline}
 \frac{2}{2+(1+2x)^2} \Big\{ -(1+2x)(\vec{a}\cdot\vec{b} + \vec{b}\cdot\vec{c} +
 \vec{c}\cdot\vec{a}) \\
 + 2x \big[ (\vec{a}\cdot\vec{n})(\vec{b}\cdot\vec{n}) + 
 (\vec{b}\cdot\vec{n})(\vec{c}\cdot\vec{n}) +
 (\vec{c}\cdot\vec{n})(\vec{a}\cdot\vec{n}) 
 \big]\Big\} \le 1.
 \label{eq:Bell_inequality_4}
 \end{multline}
In the configuration $\vec{a}+\vec{b}+\vec{c}=\vec{0}$,
$\vec{a}\cdot\vec{n} = \vec{b}\cdot\vec{n} = \vec{c}\cdot\vec{n} = 0$
(\ref{eq:Bell_inequality_4}) takes the form
 \begin{equation}
 \frac{3(1+2x)}{2+(1+2x)^2} \le 1,
 \label{eq:Bell_inequality_5}
 \end{equation}
and one can easily check that this inequality is violated for
$0<x<1/2$. (Let us note that the value $x=0$ corresponds to the
nonrelativistic limit for which the inequality is not violated).
The dependence of the left side of the inequality
(\ref{eq:Bell_inequality_5}) is shown on the Fig.~\ref{fig:Bell_1}.

\begin{figure}
\centering
\includegraphics[scale=0.70]{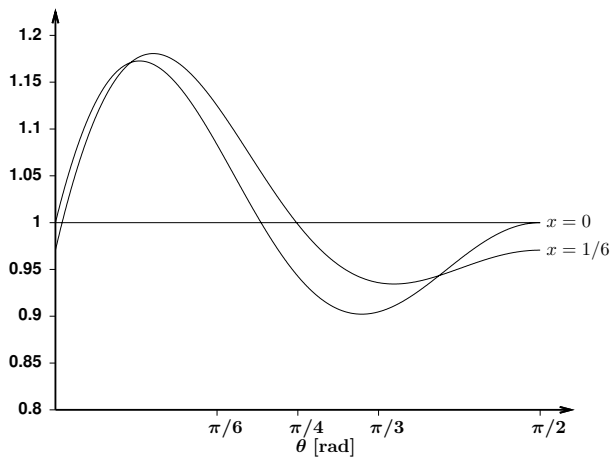}
\caption{The plot shows dependence of the 
left side of the inequality (\ref{eq:Bell_inequality_8}) on $\theta$.
The value $x=0$ corresponds to the nonrelativistic case
The plotted function has maximum value equal to $3\sqrt{2}/4$
for $x=(\sqrt{2}-1)/2$.} \label{fig:Bell_2}
\end{figure}

\begin{figure}
\centering
\includegraphics[scale=0.70]{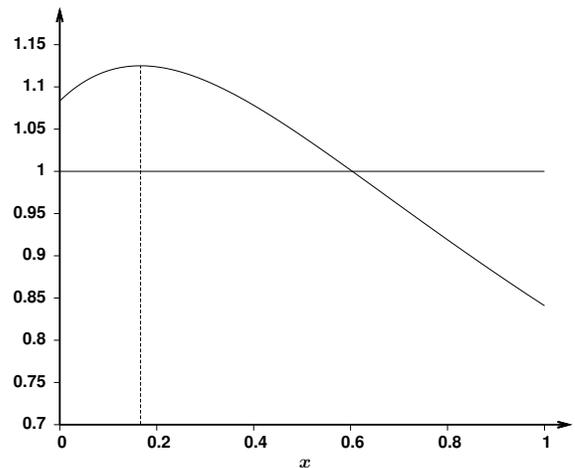}
\caption{The plot shows dependence of the 
left side of the inequality (\ref{eq:Bell_inequality_8}) on $x$ for 
$\theta=2\pi/3$ corresponding to the configuration
$\vec{a}+\vec{b}+\vec{c}=\vec{0}$. 
The plotted function has maximum value equal to $9/8$
for $x=1/6$.} \label{fig:Bell_3}
\end{figure}

In the paper \cite{cab_Mermin1980} another Bell-type inequality,
which is violated in the nonrelativistic case, is
considered. This inequality contains not only a correlation function but
also the average value of the difference of spin projections measured
by Alice and Bob and has the following form:
 \begin{equation}
 \sum_{\lambda,\sigma}|\lambda-\sigma|
 P_{\lambda\sigma}(\vec{a},\vec{b}) \ge
 C_{\bold{ac}} + C_{\bold{bc}}.
 \label{eq:Bell_inequality_6}
 \end{equation}
We have calculated the probabilities
$P_{\lambda\sigma}(\vec{a},\vec{b})$
[Eqs.~(\ref{seq:prawdopodob_jawnie_CMF})] therefore we can analyze the
inequality (\ref{eq:Bell_inequality_6}) also in the relativistic
framework. Inserting (\ref{seq:prawdopodob_jawnie_CMF}) and 
(\ref{eq:f_korelacji_jawna_nierel})
into (\ref{eq:Bell_inequality_6}) we obtain the inequality
\begin{multline}
 \frac{2}{2+(1+2x)^2} \Big\{ -(1+2x)(\vec{a}\cdot\vec{b} + 
 \vec{b}\cdot\vec{c} +
 \vec{c}\cdot\vec{a}) \\
 + 2x \big[ (\vec{a}\cdot\vec{n})(\vec{b}\cdot\vec{n}) + 
 (\vec{b}\cdot\vec{n})(\vec{c}\cdot\vec{n}) +
 (\vec{c}\cdot\vec{n})(\vec{a}\cdot\vec{n}) 
 \big] \\
 + \frac{1}{2} \big[ 
 (\vec{a}\cdot\vec{b})+2x(\vec{a}\cdot\vec{n})(\vec{b}\cdot\vec{n}) 
 \big]^2
 \Big\} \le 1.
 \label{eq:Bell_inequality_7}
 \end{multline}
This inequality is stronger than (\ref{eq:Bell_inequality_4}). 
Let us analyze the inequality (\ref{eq:Bell_inequality_7}) in the 
configuration considered in \cite{cab_Mermin1980}, that is let us
assume that $\vec{a}$, $\vec{b}$, $\vec{c}$ are coplanar and
$\vec{a}\cdot\vec{b}=\cos{(\pi-2\theta)}$,
$\vec{a}\cdot\vec{c}=\vec{b}\cdot\vec{c}=\cos(\pi/2+\theta)$. Moreover
let us assume that 
$\vec{a}\cdot\vec{n} = \vec{b}\cdot\vec{n} = \vec{c}\cdot\vec{n} =
0$. In this configuration (\ref{eq:Bell_inequality_7}) takes the
following form
 \begin{equation}
 \frac{2(1+2x)\big[ 2\sin{\theta} + \cos(2\theta) \big] + 
 \cos^2(2\theta)}{2+(1+2x)^2} \le 1.
 \label{eq:Bell_inequality_8}
 \end{equation}
We have shown the dependence of the left side of (\ref{eq:Bell_inequality_8}) 
on $\theta$ for two chosen vales of $x$: $x=0$
corresponding to the nonrelativistic case and $x=1/6$
on Fig.~\ref{fig:Bell_2}.
We show the dependence of the left side of
(\ref{eq:Bell_inequality_8}) on $x$ 
in Fig.~(\ref{fig:Bell_3}).
We have chosen $\theta=2\pi/3$ corresponding to the configuration
$\vec{a}+\vec{b}+\vec{c}=\vec{0}$ considered earlier. 

Summarizing, our analysis shows that relativistic vector bosons
violate Bell inequalities more than nonrelativistic spin-1
particles and that the degree of violation of the Bell inequality
depends on the particle momentum.

\section{Conclusions}

We have discussed joint probabilities and the correlation function of
two relativistic vector bosons in the framework of quantum field
theory. We have classified two-particle covariant states and defined
the observables corresponding to detectors measuring the spin of the
particles with momenta belonging to a given region of momentum
space. Using this formalism we have explicitly calculated the
correlation function and the probabilities in the scalar state. We
observed strange behavior of the correlation function and the
probabilities. It appears that in the CMF frame for the definite
configuration of the particles momenta and directions of the spin
projection measurements, the correlation function still depends on
the value of the particles momenta. Recall that for two fermions the
correlation function in CMF frame in the singlet state does not depend
on momentum \cite{CR2006}. Furthermore, in the bosonic case for fixed
spin measurement directions, the correlation function (and the
probabilities) can have extrema for some finite values of the
particles momenta. This affects the degree of violation of
Bell-type inequalities. We have discussed the Bell-type inequality
(\ref{eq:Bell_inequality_1})
which is fulfilled in the nonrelativistic limit but is violated in
some finite region of the particles momenta.
We have also shown that Bell-type inequality (\ref{eq:Bell_inequality_6})
 which is violated for
nonrelativistic spin-1 particles in the relativistic case
is violated more in
some finite region of the particles momenta.

\begin{acknowledgments}
This work has been supported by the University of Lodz grant and
partially supported by the European Social Fund and Budget of State
implemented under the Integrated Regional Operational
Programme. Project: GRRI-D.
\end{acknowledgments}

\appendix
\section{Explicit form of matrices $N^{\alpha\beta}$, $M^{\alpha\beta}$ and
$T^{\alpha\beta}$} The explicit form of matrices
(\ref{eq:nowe_macierze_nmt}) is
\begin{widetext}
\begin{subequations}
\label{seq:nowe_macierze_jawnie}
    \begin{equation}
    N^{\alpha\beta}(\bold{q},\boldsymbol{\omega}) = 
                  \left(
                        \begin{array}{c|c}
                          0 & \tfrac{i}{m}(\bold{q}\times\boldsymbol{\omega})^{\rm{T}} \\\hline
                          -\tfrac{i}{m}(\bold{q}\times\boldsymbol{\omega}) & -i\epsilon^{ijk}\omega^k+
                          i\left[\tfrac{\bold{q}\otimes(\bold{q}\times\boldsymbol{\omega})^{\rm{T}}-
                          (\bold{q}\times\boldsymbol{\omega})\otimes\bold{q}^{\rm{T}}}
                          {m(m+q^0)}\right]^{ij} \\
                        \end{array}
                      \right),    
   \end{equation}
   \begin{multline}
    M^{\alpha\beta}(\bold{q},\boldsymbol{\omega}) \\
    = \left(
      \begin{array}{c|c}
        \tfrac{\bold{q}^2-(\boldsymbol{\omega}\cdot\bold{q})^2}{m^2} & \tfrac{\vec{q}^{\rm{T}}}{m}
        \big[\tfrac{q^0}{m}-\tfrac{(\boldsymbol{\omega}\cdot\vec{q})^2}{m(m+q^0)}\big]-
        \boldsymbol{\omega}^{\rm{T}}\tfrac{\boldsymbol{\omega}\cdot\vec{q}}{m} \\\hline
        \tfrac{\vec{q}}{m}
        \big[\tfrac{q^0}{m}-\tfrac{(\boldsymbol{\omega}\cdot\vec{q})^2}{m(m+q^0)}\big]
        - \boldsymbol{\omega}\tfrac{\boldsymbol{\omega}\cdot\vec{q}}{m}
        &
        {\id} - \boldsymbol{\omega}\otimes\boldsymbol{\omega}^{\rm{T}} -
        \tfrac{\boldsymbol{\omega}\cdot\vec{q}}{m(m+q^0)}
        \big[\boldsymbol{\omega}\otimes\bold{q}^{\rm{T}} +
        \bold{q}\otimes\boldsymbol{\omega}^{\rm{T}}\big] +
         \big[1 - \tfrac{(\boldsymbol{\omega}\cdot\vec{q})^2}{(m+q^0)^2}\big]
        \tfrac{\bold{q}\otimes\bold{q}^{\rm{T}}}{m^2}\\
              \end{array}
    \right),
   \end{multline}
   \begin{equation}
    T^{\alpha\beta}(\bold{q},\boldsymbol{\omega}) 
    = \left(
      \begin{array}{c|c}
        \tfrac{(\boldsymbol{\omega}\cdot\vec{q})^2}{m^2} &
        \tfrac{\boldsymbol{\omega}\cdot\vec{q}}{m}
           \big[\boldsymbol{\omega}^{\rm{T}} +
        \tfrac{(\boldsymbol{\omega}\cdot\vec{q})\bold{q}^{\rm{T}}}{m(m+q^0)}\big] \\\hline
        \tfrac{\boldsymbol{\omega}\cdot\vec{q}}{m} \big[\boldsymbol{\omega}+
        \tfrac{(\boldsymbol{\omega}\cdot\vec{q})\vec{q}}{m(m+q^0)}\big] &
        \boldsymbol{\omega}\otimes\boldsymbol{\omega}^{\rm{T}}+
        \tfrac{\boldsymbol{\omega}\cdot\vec{q}}{m(m+q^0)}
        \big[\boldsymbol{\omega}\otimes\bold{q}^{\rm{T}}+
        \bold{q}\otimes\boldsymbol{\omega}^{\rm{T}}\big]
        +\tfrac{(\boldsymbol{\omega}\cdot\vec{q})^2}{m^2(m+q^0)^2}\vec{q}\otimes\vec{q}^{\rm{T}}
      \end{array}
    \right).
    \end{equation}
\end{subequations}
\end{widetext}

\end{document}